# Laser-Induced Vibrational Frequency Shift

D. L. Andrews and R. G. Crisp

*Nanostructures and Photomolecular Systems, School of Chemical Sciences, University of East Anglia, Norwich, NR4 7TJ, U.K.*

(Dated: 25 February 2005)

Abstract

A mechanism is explored whereby intense laser radiation induces an optical force between the constituent atoms of a molecule. In the case of a diatomic molecule the effect results in a modification of the vibrational potential, and using perturbation theory it is shown that this reduces the stretching frequency. Model calculations on selected diatomics indicate that the extent of the frequency shift should, under suitable conditions, be detectable by Raman spectroscopy.

It is a commonly exploited virtue of laser spectroscopy that the incident radiation does not in general physically modify the sample. However, it transpires that the application of intense optical radiation can induce a force between neighboring atoms, modifying their chemical bonding and producing measurable changes in the vibrational spectrum. Specifically an energy shift results from the engagement of two atoms, coupled by a dynamic dipolar interaction, in collective forward scattering of laser light. Through detailed analysis of a diatomic system, the aim of this Letter is to demonstrate that the associated force, induced by the presence of intense optical radiation, is attractive. Moreover it can result in a shift of the vibrational frequency that is above the limit of experimental detection.

Cooperative (pairwise) stimulated scattering of light is a process that operates for any pair of bonded atoms, AB, within a larger molecule. For simplicity the theory of the corresponding optically induced force is to be developed here specifically for a diatomic molecule. In order to develop tractable expressions for this system, it is expedient to treat the force-generating mechanism in terms of optical interactions with the individual atoms of the diatomic – justified on the basis that electrons involved in bonding cannot contribute to such a mechanism. Since the quantum electrodynamical (QED) description of electromagnetic coupling with the radiation field includes a summation over all charges in the molecule, the non-bonding charge interactions are separable into two terms, one describing the interaction of atom A with the radiation field, the other the interaction of atom B. Fourth-order perturbation theory generates the dominant contribution to the mechanism; overall there are 24 time-orderings, one of which is shown in Fig. 1. Here, the annihilation of a photon from the laser radiation mode at one atom promotes an interatomic transfer of energy resulting in the emission of a photon, by the second atom, back into the throughput radiation mode. It is of interest to note that this mechanism is applicable to any pair of polarisable particles [1-8]. In the present context, the separation of the constituent particles is sufficiently small to allow the retarded electric dipole/electric dipole interaction potential to be represented by its near-zone limit. As shown in detail elsewhere [8, 9], the energy associated with stimulated forward scattering by the atom pair, assuming the molecule tumbles freely in the throughput laser beam, is;

$$\Delta E = -\frac{4\pi I \alpha'_A \alpha'_B}{cR^3}, \qquad (1)$$

in which $\alpha'_A$ and $\alpha'_B$ are the volume polarizabilities of the two atoms, $I$ is the irradiance of the applied laser light and $R$ the atomic separation. The associated classical force ($-d\Delta E/dR$) also acquires a negative sign, thereby signifying an attractive interaction. Quantum mechanically, the potential represented by equation (1) can be considered as effecting perturbations on the wavefunctions of a simple harmonic oscillator.

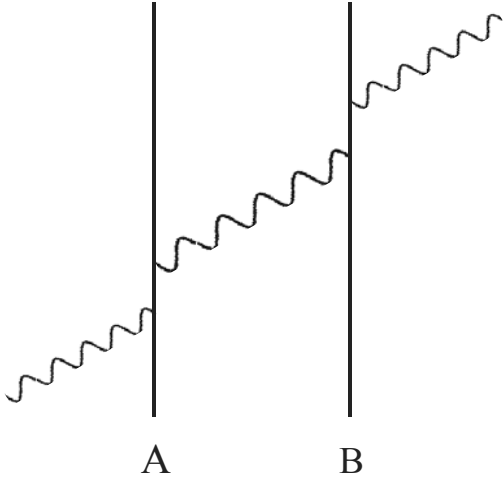

FIG. 1. Typical time-ordered diagram for laser-induced energy shift (time progressing upwards), here showing the annihilation and subsequent creation of a laser photon at atoms A and B, mediated by virtual photon transfer.

The unperturbed vibrational states of the diatomic molecule are represented by states labeled $|n\rangle$, and spectroscopic interest mainly focuses on the fundamental transition $0 \to 1$ for which in the absence of radiation $\Delta E_0 = \hbar\omega$, where $\omega$ is the circular frequency of oscillation. The Hamiltonian describing vibrations of the diatomic in the presence of laser radiation is modified by the inclusion of the irradiance-dependent correction, $\hat{H}_{int}$, from equation (1), and expressible as;

$$\hat{H}_{int} = -\frac{4\pi I \alpha'_A \alpha'_B}{cr_0^3}\left(1+\frac{x}{r_0}\right)^{-3} = \kappa\left(1+\frac{x}{r_0}\right)^{-3}, \quad (2)$$

where $r_0$ is the equilibrium bond length, $x = i(\hbar/2\mu\omega)^{\frac{1}{2}}(a-a^\dagger)$ is an operator for the displacement from equilibrium with $\mu = m_A m_B/(m_A + m_B)$ the reduced mass; $m_A$ and $m_B$ are the masses of the two atoms. The structure of the operator $x$ is defined in terms of raising and lowering operators [10], $a^\dagger$ and $a$ respectively, operating on the basis states as: $a^\dagger|n\rangle = \sqrt{n+1}|n+1\rangle$ and $a|n\rangle = \sqrt{n}|n-1\rangle$. The laser-perturbed energy associated with vibrational state $|n\rangle$ can be secured from the leading terms in a perturbation series;

$$E^{(n)} \simeq E_0^{(n)} + \langle n|\hat{H}_{int}|n\rangle + \sum_r \frac{\langle n|\hat{H}_{int}|r\rangle\langle r|\hat{H}_{int}|n\rangle}{E_n - E_r}, \quad (3)$$

terms of higher order contributing negligibly. With a Maclaurin series expansion of (2) substituted into (3), discounting any terms in $x^3$ or higher powers, the result is as follows;

$$E^{(n)} = E_0^{(n)} + \kappa\left[\langle n|n\rangle - \frac{3i\sqrt{\hbar}}{(2\mu\omega)^{\frac{1}{2}}r_0}\langle n|(a-a^\dagger)|n\rangle - \frac{6\hbar}{2\mu\omega r_0^2}\langle n|(a^2 - aa^\dagger - a^\dagger a + a^{\dagger 2})|n\rangle\right]$$
$$-\frac{9\kappa^2\hbar}{2\mu\omega r_0^2}\left[\frac{\langle n|(a-a^\dagger)|n-1\rangle\langle n-1|(a-a^\dagger)|n\rangle}{E_n - E_{(n-1)}} + \frac{\langle n|(a-a^\dagger)|n+1\rangle\langle n+1|(a-a^\dagger)|n\rangle}{E_n - E_{(n+1)}}\right]. \quad (4)$$

With $\langle m|n\rangle = \delta_{mn}$ and $E_n = (n+\tfrac{1}{2})\hbar\omega$ it transpires that the vibrational energy uptake associated with the fundamental transition is;

$$E^{(1)} - E^{(0)} = E_0^{(1)} - E_0^{(0)} + \frac{6\kappa\hbar}{\mu\omega r_0^2} = \hbar\omega - \frac{24\pi\hbar I \alpha'_A \alpha'_B}{c\mu\omega r_0^5} \quad (5)$$

Thus the vibrational wavenumber of the molecule, as interpreted from spectroscopic measurement of the fundamental transition under these conditions, is shifted to a lower value, the shift given by;

$$\Delta\bar{\nu} = \bar{\nu}' - \bar{\nu} = -12 I \alpha'_A \alpha'_B / c^2 \mu\omega r_0^5. \quad (6)$$

A corresponding perturbation theoretic evaluation of the corrected wavefunctions associated with the diatomic vibrations allows evaluation (through the expectation value of $x$) of a corresponding shift in the equilibrium bond length;

$$r_0' - r_0 = \frac{3\kappa}{\mu\omega^2 r_0} - \frac{9\kappa^2(n^2+n+1)}{\mu^2\omega^4 r_0^3}, \quad (7)$$

the magnitude depending on the vibrational quantum number $n$. Since $\kappa$ is negative, this result invariably signifies a radiation-induced reduction in bond length even for $n = 0$, consistent with the attractive nature of a corresponding classical force. Although, intuitively, an attractive force would be expected to increase the vibrational frequency, Fig. 2 shows how a modification of the simple harmonic potential well, through the laser-induced interaction potential, forms a new and slightly broader potential well with quantum levels of correspondingly slightly lower energy.

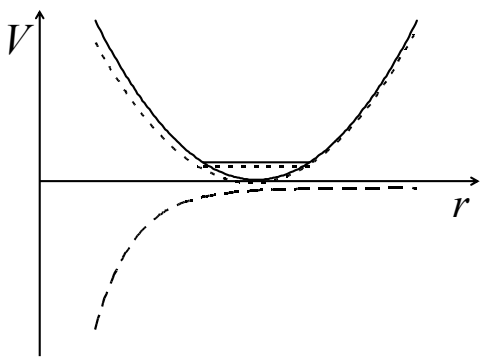

FIG. 2. Schematic illustration of the modification to a simple harmonic potential (solid line) by a laser-induced interatomic potential (dashed line), forming a shifted potential well (dotted line). Note the quantum vibrational level is lowered by the perturbation.

Before proceeding to an evaluation based on the above theory, certain assumptions in the model calculations invite comment. First, although the reduction in equilibrium bond length has not been taken into account in evaluating frequency shifts; calculations based on equation (7) suggest that the ground state bond length is typically reduced by less than 0.01% of its unperturbed value, even at an applied irradiance of $10^{15}$ W m$^{-2}$. Secondly, whereas the dynamic polarizabilities depend on the frequency of the radiation field, static polarizabilities are more readily available and have been used in the calculations. While using the free atom polarizabilities is a good approximation, especially for large atoms, a more precise result would be obtainable by summing the coupling of the electromagnetic field to each individual charge in the molecule; this avenue of development is currently under investigation.

Table I gives calculated values for the wavenumber shifts, $\Delta \bar{\nu}$, of several diatomic molecules selected on the basis of a trade-off between polarizability and reduced mass. The atomic and molecular data was obtained from reference materials [11, 12], and a pulse intensity of $10^{15}$ W m$^{-2}$ was chosen – a level readily achievable from a standard femtosecond laser, but not high enough to produce dissociation. In passing we note that the use of pulsed laser light in no way undermines the phenomenon of interest, because the vibrational frequency measurements are made only while the light is present. Also, even when such pulses are short compared with the vibrational period, the usual Born-Oppenheimer rules apply, i.e. nuclear motion responds essentially instantaneously to the (laser-modified) electronic environment.

Raman spectroscopy appears to afford the ideal technique for experimental verification of the vibrational shifts predicted under high-irradiance conditions. Both homonuclear and heteronuclear diatomic vibrations are Raman-active, and therefore it is possible for the same laser to be used both to induce the frequency shift and to measure it. With current dispersive instrumentation and high-resolution measurement techniques, it is practicable to achieve a resolution of 0.04 cm$^{-1}$ [13]. On comparison with the results given in Table I, it thus transpires that although the wavenumber shift will be too small to detect in the case of the diatomic halogens; the shifts for hydrogen halides are at or above the detection threshold, and in the case of hydrogen the shift should be readily measurable.

In conclusion we observe that the use of ultrashort laser pulses signifies pulse durations which may in some cases prove to be similar in magnitude to the vibrational period, 100 fs corresponding to a vibrational wavenumber of 300 cm$^{-1}$ for example. There is scope to explore in future work the dynamical aspects of the laser-induced frequency shift that has been identified.

## Acknowledgments

We would like to thank D S Bradshaw for helpful comments on the manuscript, and Dr U A Jayasooriya for encouraging comments on the Raman aspects of our work. RGC thanks the EPSRC for a studentship, during which this work was carried out.

| Molecule | $\bar{\nu}$ (cm$^{-1}$) | $\bar{\nu}'$ (cm$^{-1}$) | $\Delta\bar{\nu}$ (cm$^{-1}$) |
|---|---|---|---|
| I$_2$ | 214.47 | 214.46 | -0.01 |
| Br$_2$ | 323.27 | 323.26 | -0.01 |
| Cl$_2$ | 556.11 | 556.10 | -0.01 |
| HI | 2309.29 | 2309.20 | -0.09 |
| HBr | 2649.04 | 2648.98 | -0.06 |
| HCl | 2988.80 | 2988.74 | -0.06 |
| HF | 4140.79 | 4140.73 | -0.06 |
| H$_2$ | 4402.41 | 4402.02 | -0.39 |

TABLE I. For selected diatomic molecules, wavenumbers of the fundamental vibrational transition in the absence ($\bar{\nu}$) and presence ($\bar{\nu}'$) of laser light at an irradiance 10$^{15}$ W m$^{-2}$, with the associated shift $\Delta\bar{\nu}$.